# Photo and acoustic emissions from the non-equilibrium phase transition at the interface during cavitation

Shamit Shrivastava and Robin O. Cleveland

*Department of Engineering Science, University of Oxford*

Draft: 21 November 2019

**Abstract**

This study investigates the emission of light and sound from cavitation caused by intense pressure pulses in water. Based on time-resolved measurements of (a) pressure waveform at the focus, (b) light scattering upon cavitation, (c) acoustic emission, and (d) photoemission (sonoluminescence) it is shown that emissions occur upon the creation or expansion as well as the collapse of the cavity. These results suggest that the thermodynamic irreversibility, resulting from non-equilibrium phase transition and changes in surface entropy, is a basis for photo and acoustic emissions during cavitation.

**Introduction**

The thermodynamics of phase change or phase transitions plays a fundamental role in a wide range of phenomena[1]. Thermodynamic processes are typically studied under the assumption of reversibility and equilibrium, attained via a quasi-static process, i.e., when the time scales of the experiment are much longer than the relaxation timescales. However, due to the involvement of critical phenomena, such as critical slowing down[2], phase transitions are more likely to occur under non-equilibrium and irreversible conditions[3]. Such conditions are expected to produce entropy in the form of acoustic and/or photo emissions. Indeed, acoustic emissions have been investigated in a variety of materials undergoing phase transitions[4]. These transitions are accompanied by large changes in volumes, and the emissions are typically stronger during contraction phase than expansion phase. Thus acoustic and photo emissions are key observables of a phase transition phenomena in the framework of irreversible thermodynamics. Here we apply this framework to understand the nature of acoustic and photo emissions that accompany water to vapour transition during acoustic cavitation.

Acoustic cavitation or vaporisation of a fluid under dynamic tensile stress and the subsequent collapse of the cavity represents a far from equilibrium phase transition phenomenon. The phenomenon is accompanied by the emissions of sound and sometimes light[5]. As discussed, these emissions can provide insights into the mechanisms of dissipation during dynamic phase transitions in general[3,4]. In particular, the magnitude of these emissions is expected to be directly related to the rate of non-equilibrium heat transfer during phase transition.

Recently we provided evidence that the entropy or the heat content of the interface plays a critical role in cavitation dynamics. We showed that during the expansion of a cavitation cluster, sudden transfer of latent heat of vaporisation across the interface can condense the surfactants at the interface[6]. The entropy of the interface is given by $\Delta S_i = -\frac{d\sigma(T)}{dT}\Delta A$ where $\sigma$ is the surface tension $A$ is the surface area and $T$ is the temperature of the interface[7]. Therefore, during acoustic cavitation even when there is no vaporization, heat transfer is required to allow a change in surface area; to the surface during expansion and from the surface during collapse.

Furthermore, as the velocity of the interface begins to approach the velocity of sound in the surrounding media, there will be non-equilibrium heat transfer, resulting in dissipation via acoustic and photo emissions. Therefore, acoustic and photo emissions should be expected not only during collapse of the cavity or compression, which is well known but also during the expansion of the cavity or vaporisation as observed in other materials[3]. Also, the emissions should be directly related to the rate of expansion and compressions (collapse).

Compared to emissions from oscillating bubbles where emission peaks are observed once per cycle, emissions from an acoustic impulse induced cavitation shows a signature double peak[5]. The first emission coincides with the arrival of the pressure impulse at the focus. There is then a period in which the cavity grows and collapses inertially, at which point a second emission is observed. Depending on the magnitude of dissipation during the process, the collapse may be followed by a rebound and a collapse again[8]. A general assumption in the community has been that the first emission results from the crushing of the pre-existing bubbles at the focus by the acoustic source.

This study provides experimental evidence that both photo and acoustic emission of the first peak occur during the negative phase of the pressure impulse or from the expansion of the water to vapour phase, which is in line with the observations in other materials[4]. Furthermore, as expected from an irreversible thermodynamics perspective, the emissions are directly related to the expansion rate of the cavity. The study thus strongly suggests that the photo and acoustic emissions during cavitation represent energy dissipation at the moving phase boundary, where non-equilibrium heat transfer occurs upon phase transition. Therefore, the propagating water-vapour interface is an important source of radiation itself, which is created during first emission peak and is annihilated during the second emission peak.

**Material and Methods**

Pressure generated by a Swiss Piezoclast® (EMS Electro Medical Systems S.A, Switzerland) were fired in to a water tank through a MYLAR membrane on one of the side panes (Fig 1). The pressure waveforms generated by the Piezoclast were measured with a PVDF needle hydrophone (Muller-Platte needle probe, Dr. Muller instruments, Oberursel, Germany) with a manufacturer specified sensitivity of 12.5mV/MPa (0.3 to 11 MHz). The needle was placed at the focus of the Piezoclast. The needle probe was removed after characterising the pressure impulse at the focus.

Acoustic emissions were monitored by using an immersed focused ultrasound transducer (U8421032, Olympus, Japan) as a passive cavitation detector (PCD) with 15 MHz central frequency, 1.6 cm diameter, and focal distance of 7.6 cm. The monitoring transducer was placed perpendicular to the proagation path on a rim of diameter 15.24 cm, concentric with the focus of the Piezoclast. The central frequency of 15MHz implies an acoustic focal region (-6 dB beam diameter) of the order of 0.06 cm ($600 \mu m$). The PCD was placed in the rim, but not all the way through to shield from directly incident pressure waves and was pre-focussed by 0.3 cm, i.e., the distance between the PCD surface and the Piezoclast focus was 7.9 cm. The acoustic emissions were quantified in terms of the root mean squared (RMS) value.

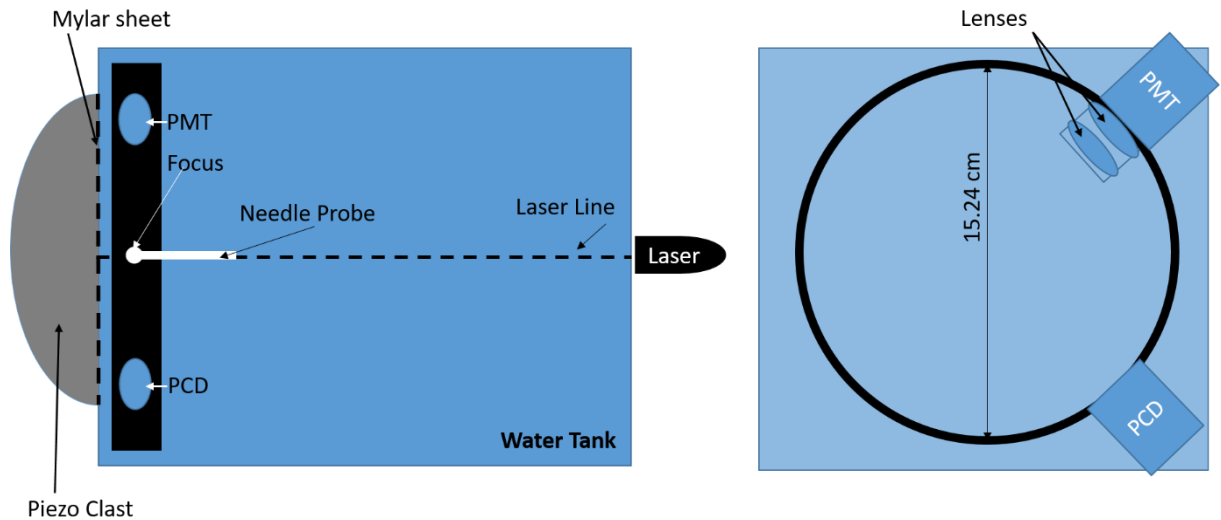

*Figure 1. Experimental Setup. On the left hand side is the side view of the setup. The pressure impulse is fired by the Piezoclast through a mylar sheet into the water tank. A passive cavitation detector (PCD) measures acoustic emissions, and a photon multiplier tube (PMT) measures the photo emissions. Coaxial with the propagation path (dashed line inside the tank) is a circular rim that holds the detectors that are perpendicular to the path of propagation. This arrangement is also shown as an along-axis view (right). The focus of the detectors is aligned with the focus of the Piezoclast. The pressure waveform at the focus was measured using a needle hydrophone (shown in white in the centre figure), which was removed during the emission measurements. For the light scattering experiments an LED laser line was aligned with the propagation path entering the water tank from the right (along the dashed line). It was to ensure that the timing of the start of the cavitation is not influenced artificially by the path of light.*

Photoemission was measured simultaneously by a photomultiplier tube (PMT, HT493-003, Hamamatsu, Japan) also placed on the circular rim, with a signal bandwidth of 8MHz and bias voltage of 0.5V. Photoemissions were quantified as the sum of output on the PMT greater than the threshold voltage of 4mV. To improve the photo collection efficiency of the PMT, two aspheric condenser lenses of focal length 1.6 cm (ACL25416U-A, Thorlabs, USA) were placed in series in front of the PMT, focusses at the Piezoclast focus.

The light scattering measurements were performed using the same PMT settings using an LED Laser with a central wavelength of 450nm (CPS450 Thorlabs, USA) that entered the water tank from the side opposite to the Piezoclast alight with the propagation path of the pressure impulse and passing through the focus. Data was acquired, digitised, and analysed using Picoscope 5444b USB Oscilloscope (Pico Technology, Cambridgeshire, UK) and Labview (National Instruments, Austin, USA).

**Results**

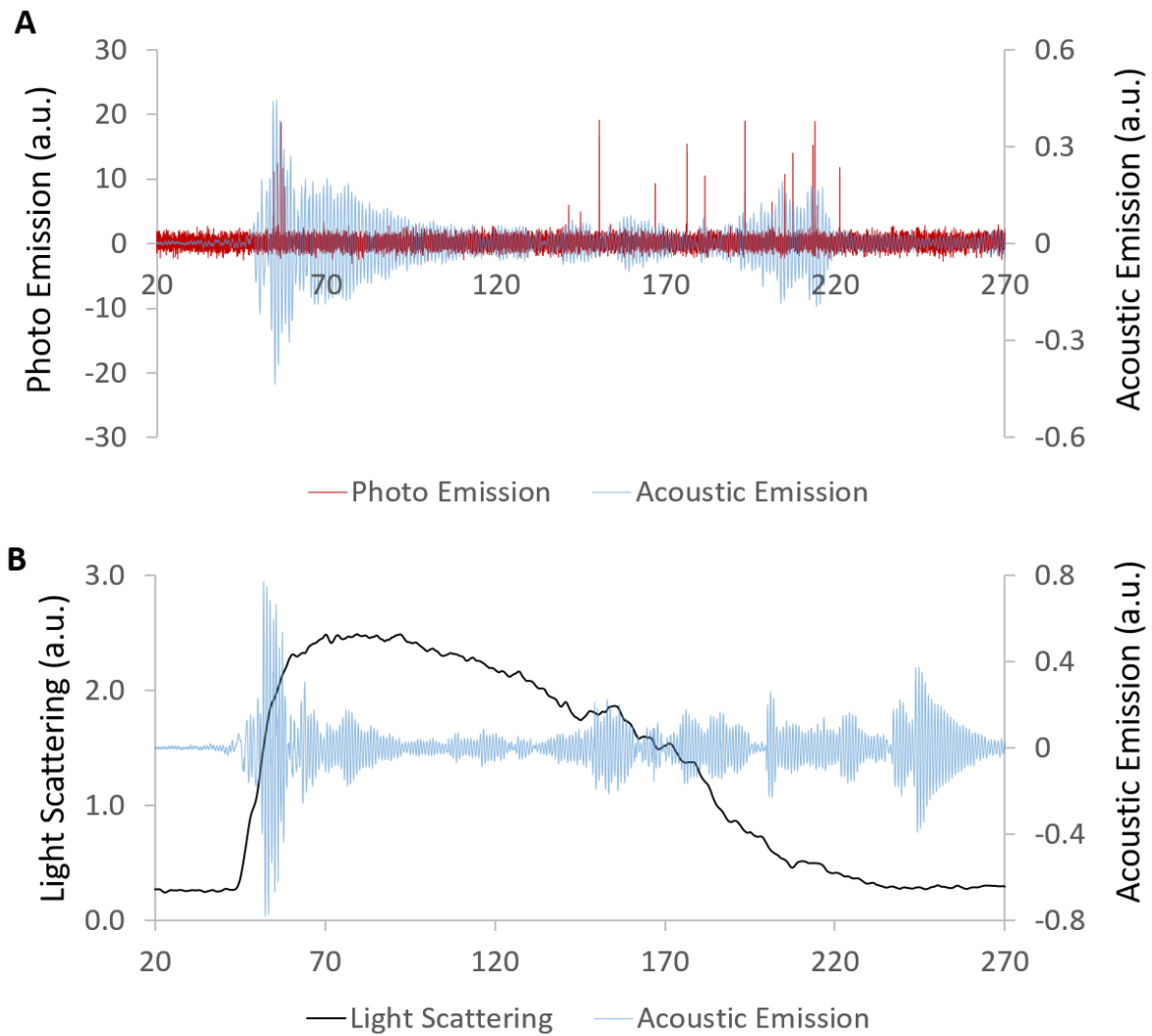

*Figure 2. Timing of events during a pressure impulse. (A) Photo and acoustic emission measured simultaneously during a pressure impulse show correlation in the emissions. The time axis for acoustic emission was shifted by $t=-52.9\mu s$, to account for the travel path of the acoustic emission from the focus to the transducer. (B) Light scattering and acoustic emission measured simultaneously during a pressure impulse (different experiment from (A)).*

Figure 2 shows representative photo and acoustic emission recorded during a pressure impulse that results in cavitation. The first photo-emissions from the focus are observed around $50\mu s$. Acoustic emissions measured simultaneously are plotted after correcting for the time of flight ($t=-52.9\mu s$) for the acoustic emission from the Piezoclast focus to the PCD. The first acoustic emissions appear in sync with the photo emission. These emissions are followed by a sequence of weaker emissions, which, as discussed below, indicate the successive collapse of cavities in the cluster, starting at $150\mu s$.

In a different experiment, light scattering from a cavitation event was recorded simultaneously with acoustic emissions (fig. 2B). An increase in scattered light indicates increase in the size of the cavity cluster at the focus. Therefore, the acoustic emissions peak appears to occur during the expansion phase of the cavitation cluster. However, fig. 2 only demonstrates the typical timing of emissions during a single pressure pulse and the statistical nature of the event needs to be ascertained.

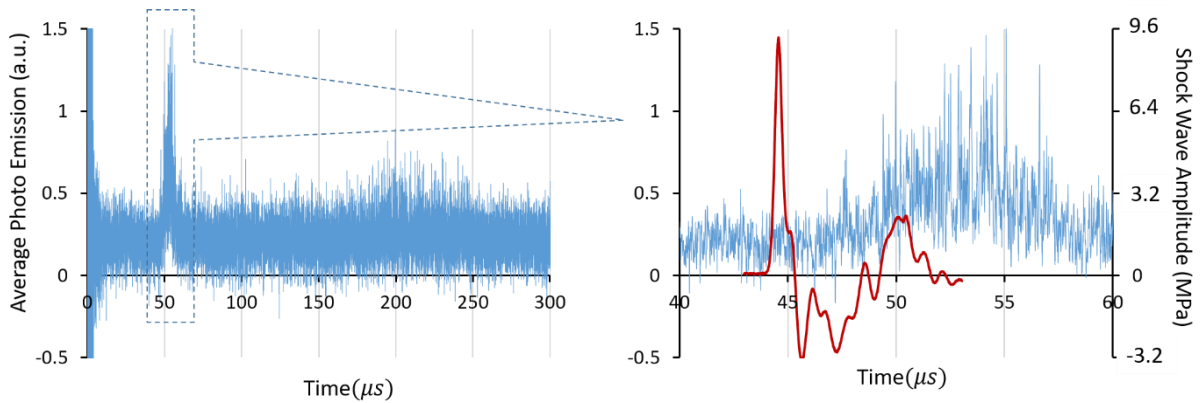

*Figure 3 Timing of photo emissions with respect to the pressure impulse. The figure on the left shows total time-resolved photoemission recorded by the PMT during 80 pressure impulses in water. The piezoclast is fired at t=0 μs, resulting in an artefact in the PMT signal. The shockwave arrives at the focus around t=44 μs, which is shown on the zoomed in figure on the right. The peak photoemission is observed around t=53 μs. Finally, weaker and temporally dispersed photo emissions are observed around t=200 μs*

To understand the origin of the initial photo-emission, the shock waveforms arriving at the focus were measured using a needle probe, as shown in fig. 1. Figure 3 plots the pressure impulse measured using the needle hydrophone at the Piezoclast focus. The waveforms exhibit a leading positive pressure of the order of 9 MPa, followed by negative tail that is not properly captured because of pressure release upon cavitation as well as limited bandwidth of the probe. The standard deviation of the arrival of peak positive pressure for 10 shock waves was $0.005\mu s$. After removing the needle probe from the focus, photo-emission was measured for 30 such shock waves with the same temporal resolution and the total emission has been plotted on the secondary axis. The experiments clearly show that the photo-emission mainly occurs during the negative pressure phase with most emission occurring for $t>50\mu s$.

Based on figure 2 and 3, emissions observed around the time of the arrival of the pressure impulse were classified as emission related to initial expansions phase. The expansion phase is assumed to start $10\mu s$ before the arrival of the shock impulse (t = 33 $\mu s$ ) and up to the peak of the scattering signal around t = 73 $\mu s$ (figure 2B). The emission observed beyond 73 $\mu s$ were classified as those from collapsing phase.

For the initial emissions, the mean peak photo emission time was observed to be 54.8 $\mu s$, with a standard deviation of 3.2 $\mu s$. The mean peak acoustic emission time was 55.1 $\mu s$ (108 $\mu s$ before path correction) with a standard deviation of 5.5 $\mu s$. The mean pair wise time difference between the two emission peaks (calculated for each experiment) was 53.2 $\mu s$ with a standard deviation of 4.3 $\mu s$. The timing is comparable to the travel time from the Piezoclast focus to the PCD plane, i.e. $t=52.9 \mu s$. The expansion rate of the bubble cluster during the initial emissions can be estimated using the time derivative of the light scattering signal. The peak expansion rate was observed at mean time of 51.5 $\mu s$ with a standard deviation of 1.5 $\mu s$.

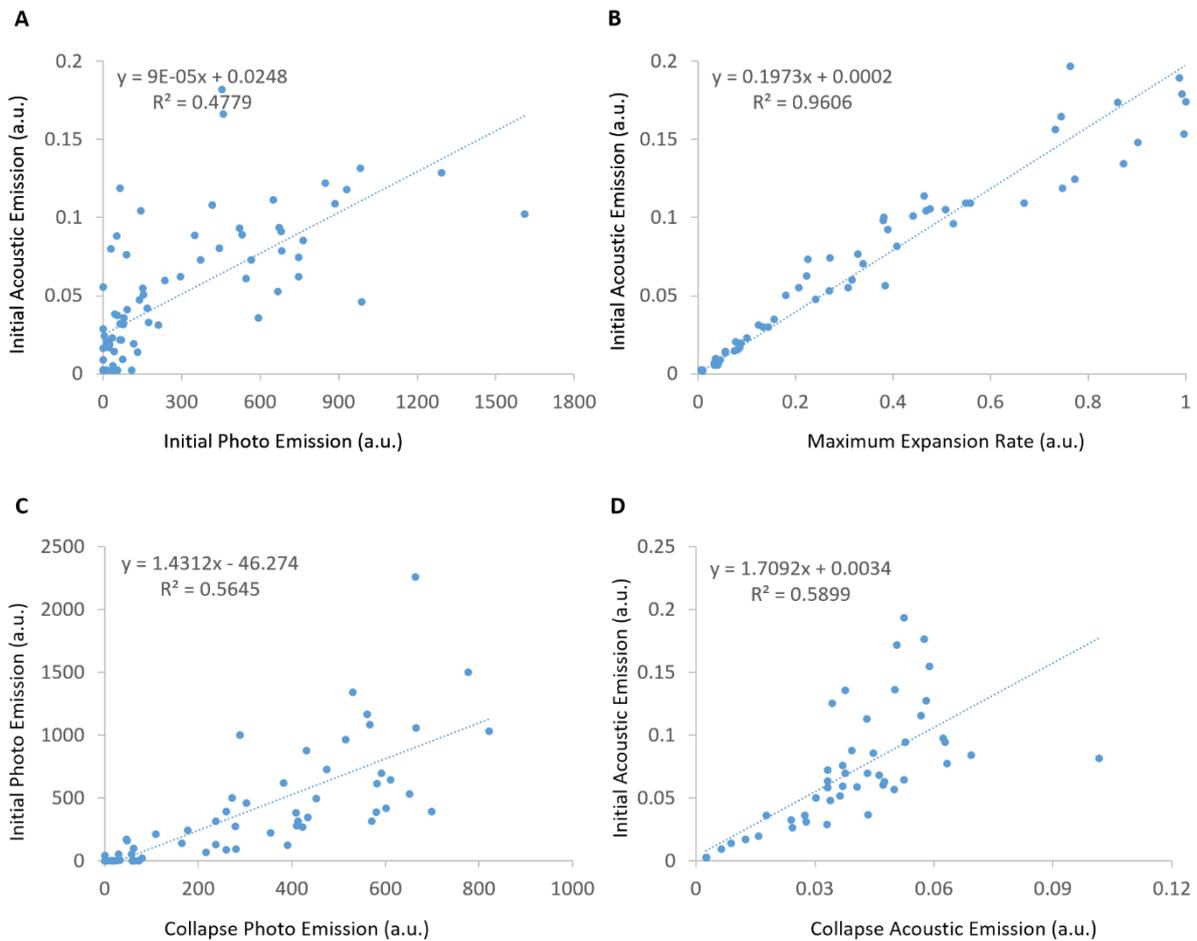

*Figure 4 Observed correlations: (A) initial acoustic emission and photoemission, (B) acoustic emission and maximum expansions rate as estimated from time derivative of the scattered light, (C) initial photo*

*emission and photo emissions during collapse phase, and (D) initial acoustic emissions and acoustic emissions during collapse phase.*

Figure 4 compiles the correlations between the magnitude of the simultaneously measured quantities and their variability in these experiments. Each dot on the plot represents a measurement during a single pressure impulse. Figure 4A plots the initial (from 33 to 73 $\mu s$) acoustic emission as function of total photo emission. It shows a threshold effect, i.e., there is no photoemission till an acoustic emission of 0.02 $V_{RMS,}$ and only beyond this value a significant photoemission appears. Beyond the threshold photo and acoustic emissions appear to be correlated. Figure 4B plots initial acoustic emission as a function of the maximum rate of the expansion of the cavitation cluster, based on time derivative of the light scattering signal. The strong correlation is indicative of the fact that observed emissions are linked to the expansion phase of the cavitation cluster. To see the correlation between the emission from expanding and collapsing phase, the corresponding photo, and acoustic emission have been compared in fig.4C and fig 4D. The emissions are correlated at lower intensities, beyond which emissions from collapse-phase are unable to keep up with strong expansion-phase emissions. The fact that expansions and collapse and emissions are correlated and of similar order (slope of the trend close to 1) indicates that the two processes are not independent.

**Discussion**

Following conclusion can be drawn from the experiments: (a) Initial photo and acoustic emissions result from the same physical process associated with the creation of the cavity. It is supported by correlations in both the timing and the magnitude of acoustic and photo emissions. However, photoemission has a higher threshold. (b) Acoustic emission is strongly correlated to the rate of expansion of the expansion of the cavity in both the timing and the magnitude, supporting the original hypothesis of irreversibility driven entropy production as a source of emission. (c) The emissions during the compression phase upon the arrival of the pressure pulse at the focus are negligible and occur predominantly during the negative pressure tail. (d) The initial emissions also show some correlation with the emissions during the collapse phase, indicating that the underlying events are not completely independent. While (a), (b) and (c) show that initial emissions are related to the creation and expansion of the interface during cavitation, (d) shows that these observations also have consequences for the emission mechanisms during the collapse of the cavity.

It is important to focus on the interface or the water-vapor boundary to understand the relation between expansion rate, dissipation, and hence emission from a thermodynamic perspective. An energy source needs to work against hydrodynamic pressure and surface tension, as well as provide for the heat content of the cavity and the interface, to create and expand the cavity. The total energy required to create a cavity reversibly is $\Delta E_{cavity} = \Delta W + \Delta Q_{rev} = P\Delta V + \sigma \Delta A + T\Delta S_{vap} + T\Delta S_{int}$, where $\Delta S_{int} =$ $-\left(d\sigma/dT\right) \Delta A$ [7]. Note that the heat transfer in the above equation is determined at the interface. However, the creation of a cavity during an intense pressure pulse will usually result in non-equilibrium conditions, which should invariably result in dissipation [3]. Creation of new interface will also result in flow of charges that compensates for the surface potential of the vapour – water or gas – water interface [9,10]. Therefore, if non equilibrium heat transfer is the source of emission, then emission or dissipation should scale with the rate of heat transfer, which in turn should scale with rate of change of cavity size. This dependence explains the direct correlation between rate of expansion and magnitude of emission (fig.4B).

The disutribution of emission peaks observed in this study is most likely from the presence of multi-bubble or cavitation cluseter[11]. In this study, we used shockwaves or acoustic impulses that resulted in cavitation cloud, or multi-bubble cavitation, where the cavities and bubbles would interact with each other as they form and collapse. As cavitation is a statistical phenomenon, the distribution of cavitation nuclei in the cluster is different for each shock. These nuclei expand and collapse while interacting with each other, which is one reason for the multiple emissions during a single shock and the distribution of emission peaks (in time and amplitude) across different shocks. Furthermore, stronger cavitation that result in larger bubbles would also imply more interaction between the bubbles, which explains the loss in correlation between initial and collapse phase emissions as the emission magnitudes increase. While strong initial phase photo and acoustic emissions are indicative of intensity of cavitation (fig 4B), corresponding collapse phase emissions became weak and less correlated due to bubble-bubble interaction(fig. 4C and 4D). Indeed, the intensity of multi-bubble sono-luminescence (MBSL) is known to be weak and distributed in time compared to single bubble sono-luminescence (SBSL) [12]. The strong emissions during SBSL are attributed to an intense spherical collapse, compressing the gases in the core to high pressures and temperature. On the other hand, spherical collapses are impeded by the interaction of bubbles in a cavitation cluster. The present study provides insights into the role of non-equilibrium phase change and heat transfer at the interface, which should apply to both MBSL and SBSL.

This study has provided evidence of emission of light during expansion of the cavity, which has not been explained before to the best of our knowledge. The emission of light during the collapse of the activity, on the other hand, has been studied extensively[13]. This study shows that the emissions from expansion and collapse are not completely independent (Fig.4C and 4D). Indeed, if non-equilibrium heat transfer as a result of phase transitions is a source of emission during expansion, it should also cause emissions during collapse. Dissipation would occur irrespective of the direction of heat transfer, which in the case of collapse, would be exothermic liquification of the vapour cavity.

Furthermore, similar to expansion, the magnitude of emissions during collapse should also correlate with the rate of change of the cavity size, which needs to be investigated during single bubble collapse. However, the nature of heat transfer during expansion and collapse is not symmetric. During collapse, the velocity of sound in vapour or gas, which is ahead of the collapsing front, is lower than the speed of sound in water that is behind the front. Therefore, the heat released from condensation across the front can accumulate at the collapsing front, accelerating the process of collapse via detonation[14]. This mechanism is missing during an expanding cavity as the process is both endothermic, and there is instead an increase in velocity across the wave-front (from vapour to water). Thus detonation provides a focussing mechanism that potentially explains the increased intensity of the collapse of a spherical cavity and SBSL[15]. Detonation as a result of exothermic phase transitions has been described in other systems before[14,16], the collapse of a vapour cavity has the additional constraint that the detonation front is spherical and imploding.

With regards to possible role of detonation during collapse of vapour cavity, a general solution $r(t)$ for an imploding detonation has been derived previously [17] assuming that all the quantities during the implosion depend on the dimensionless similarity parameter $\xi = \frac{r}{t\sqrt{q}}$; where $q$ is the heat released per unit mass of the detonating material. During collapse of a vapour-cavity $q$ is going to be the enthalpy of condensation from vapour to water[18]. The solution, however, diverge as $r \to 0$ and a complete solution will also require incorporating a photo radiation pressure $P = \frac{1}{3}a(kT)^4$, and radiation energy $W = a(kT)^4$ where $a = 2.23 \times 10^{49} erg^{-3} cm^{-3}$ and $k$ is the Boltzmann constant and T is the temperature of the radiation emitted as photons[19]. As discussed, this energy results from the dissipation (phase transition and relaxation) at the detonation front and contributes to increasing the width and limits the rise in pressure. A complete solution for an imploding detonation front that includes the radiation pressure of photons is outside the scope of this work.

A detonation based understanding of cavity collapse has certain advantages over the conventional solution based on the Raleigh-Plesset equations[8,20]. The hydrodynamic problem of bubble oscillation and collapse is typically solved using the Raleigh-Plesset equations combined with the equation of state of the gas in the core of the cavity. However, in the case of imploding cavities, application of Raleigh-Plesset equations close to collapse has caused some concerns due to their inapplicability to the shock like conditions at the collapse [15]. A detonation based solution, on the other hand, takes shock formation into account. There are other open questions as well, such as the possibility of a missing focussing mechanism during the collapse of a cavity [15], and concerns regarding the timing of collapse with respect to sono-luminescence have been highlighted before [21]. A detonation based mechanism provides possible resolution for these discrepancies as well, for example, this work suggests that the timing of maximum emissions should also have contributions from the maximum rate of disappearance of the interface during collapse. The mechanism can also shed light on how solutes in the liquid phase can effect sono-luminecence[13], as the source of the emissions is the interface, which is always in contact with the liquid phase and its solutes. On the other hand, the role of gases in the cavity or the vapour phase (e.g., xenon and argon[22]) also remains important, which enters the equations via partial pressures of the gases in the superheated vapour just ahead of the collapsing vapour-liquid boundary.

Finally, the thermodynamic framework only provides macroscopic insights regarding the fundamental constraints during phase change at the interface during cavitation. It does not provide the microscopic mechanisms (e.g., blackbody vs Bremsstrahlung radiation) that ultimately lead to the emission of photons and phonons [13].